Evolution of tetragonal phase in the FeSe wire fabricated by a novel chemical-transformation PIT process

Hiroki Izawa[1], Yoshikazu Mizuguchi[1], Toshinori Ozaki[2], Yoshihiko Takano[2], and Osuke Miura[1]

[1]Department of Electrical and Electronic Engineering, Tokyo Metropolitan University, Hachioji, Tokyo, 192-0397, Japan.

[2]National Institute for Materials Science, Tsukuba, Ibaraki, 305-0047, Japan.

Abstract

We fabricated superconducting FeSe wires by the chemical-transformation PIT process. The obvious correlation between annealing temperature and phase transformation was observed. Annealing above 500 °C produced wire-core transformation from hexagonal to tetragonal phase. Furthermore the hexagonal phase completely transformed into the tetragonal phase by annealing at 1000 °C. With increasing annealing temperature, the superconducting property was dramatically improved, associated with the evolution of the tetragonal phase.

1. Introduction

Fe-based superconductor is one of the candidate materials for superconducting applications, owing to the high transition temperature ($T_c$) and high upper critical field ($H_{c2}$) [1-3]. Several kinds of Fe-based superconducting wires have been fabricated using the superconducting materials of $BaFe_2As_2$, $SmFeAsO$, and $FeSe$ systems [4-10]. To date, the highest record of critical current density ($Jc$) over $10^4 A/cm^2$ at 4.2 K was achieved in the $Ba_{1-x}K_xFe_2As_2$ wire [5].

Among the Fe-based superconducting materials, FeSe has advantages for application because of the simplest crystal structure and composition [11-15]. Furthermore, the toxicity of Fe chalcogenides is relatively low compared to the FeAs-based compounds, and the anisotropy is low. In these respects, Fe-chcalcogenide superconductor is a great candidate material used in superconducting application. However, the present record of the highest $J_c$ in Fe-chalcogenide wires is 1027 A/cm$^2$ at 4.2 K [8], 10 times as low as that of FeAs-based wires. To achieve higher $J_c$ using Fe-chalcogenide superconductors, a new wire fabrication process should be created.

Recently we reported a successful fabrication of FeSe superconducting wire by an unconventional powder-in-tube (PIT) method based on chemical transformation of the wire core using an Fe sheath. Via a wire fabrication process, the wire core transformed from hexagonal FeSe$_{1+d}$ (non-superconducting) to tetragonal FeSe (superconducting) upon a diffusion of Fe from the pure Fe sheath by annealing [16]. In this article, we report a systematic study on the annealing temperature dependences of structural changes and transport properties of the FeSe superconducting wires fabricated by the chemical-transformation PIT process.

2. Experimental details

FeSe superconducting wires were fabricated by the chemical-transformation PIT process. Figure 1 shows a schematic chart for wire fabrication process. Firstly, we synthesized precursor powders of hexagonal FeSe$_{1.2}$ by solid state reaction. Pure Fe powder (99.9 %) and pure Se chips (99.999 %) were used as starting materials. These materials were sealed into an evacuated quartz tube, and heated at 700 °C for 10 h. The obtained precursor was packed into a pure Fe tube with outer and inner diameters of 6.2 and 4.0 mm, respectively. The tube sealed with two edge caps of pure Fe was groove-rolled into a rectangular wire with a size of ~2mm. The obtained wire was cut into several pieces, sealed into an evacuated quartz tube, and then annealed at various annealing temperature ($T_a$) of 400 – 1000 °C. The cross section of the wire was observed using an optical microscope. The crystal structure was characterized by x-ray diffraction (XRD) using a CuKα radiation. Temperature dependence of total resistivity down to 2 K was measured using a four-terminal method, where the total resistivity was estimated using the total cross-sectional area including Fe sheath.

3. Results and discussion

Figure 2 shows the optical-microscope images of the obtained cross section for $T_a$ = (a)400, (b)800 and (c)1000, respectively. Dense core without voids was observed for all specimens.

Figure 3 shows the XRD patterns for the $FeSe_{1.2}$ precursor and wire core annealed at 400-1000 °C. The peaks of $FeSe_{1.2}$ precursor was well indexed using the hexagonal space group of $P_{63}$/mmc. The estimated lattice parameters were $a$ = 3.602(2) Å and $c$ = 5.894(6) Å. With increasing $T_a$, peaks of the hexagonal phase were suppressed; in contrast the peaks of tetragonal phase were appeared. For $T_a$ > 500 °C, the ratio of the tetragonal phase to the hexagonal phase was over 70%. Finally, the peaks of the hexagonal phase disappeared at $T_a$ = 1000 °C, while a small peak of Fe was detected. The lattice constants of the tetragonal phase were estimated to be $a$ = 3.777(2) Å and $c$ = 5.535(7) Å. To discuss the evolution of the hexagonal-tetragonal transformation by annealing, we plotted the $T_a$ dependence of the existence ratio of the hexagonal phase in fig. 4. The ratio of the hexagonal phase was defined as $I_{hex} / ( I_{hex} + I_{tet} )$, where $I_{hex}$ and $I_{tet}$ are the first peak intensities for the hexagonal and tetragonal phase, respectively. Dramatic changes were observed at two critical points. The first critical $T_a$ exists between 400 and 500 °C. The second critical $T_a$ is between 900 and 1000 °C. This shows that the annealing temperature would be a key parameter of FeSe wire fabrication process.

Figure 5 shows the temperature dependence of total resistivity from 50 to 2 K for $T_a$ = 400 - 1000 °C. With increasing $T_a$, the drop of resistivity corresponding to superconducting transition became larger and transition became sharper. Figure 6 shows the annealing temperature dependence of $T_c^{onset}$ and $T_c^{offset}$, where the resistivity was 90 % and 10 % of normal-state resistivity, respectively. Both $T_c^{onset}$ and $T_c^{offset}$ were enhanced with increasing $T_a$. Around $T_a$ = 400 - 500 °C, a dramatic enhance of $T_c^{onset}$ was observed. For $T_a$ = 1000 °C, a large enhance of $T_c^{onset}$ and $T_c^{offset}$ were observed, and the highest critical current density of 218 A/cm$^2$ (at 4.2 K and 0 T) was obtained for this wire as reported in Ref. 16. These tendencies seem to correspond with that in Fig. 4, indicating that the annealing temperature is one of the most important parameter to enhance transport properties of FeSe wire as well. By optimization of annealing conditions, the transport properties of FeSe wire will be greatly enhanced.

## 4. Summary

We fabricated superconducting FeSe wires by the novel chemical-transformation PIT process, and investigated annealing temperature dependence of structural and superconducting properties. The obvious correlation between annealing temperature and phase transformation was observed. Annealing above 500 °C produced the wire-core transformation from hexagonal to tetragonal phase. Furthermore the hexagonal phase completely transformed into the tetragonal phase by annealing at 1000 °C. With increasing annealing temperature, the superconducting property was improved, associated with the evolution of tetragonal phase.

Acknowledgements

This work was partly supported by Grant-in-Aid for Scientific Research (KAKENHI).

Figure captions

Fig. 1. Schematic chart for fabrication of FeSe superconducting wire by the chemical-transformation PIT process. The schematic images of crystal structure were depicted using VESTA [17].

Fig. 2. Typical optical-microscope images of the obtained cross section for $T_a$ = (a)400, (b)800, and (c)1000 °C.

Fig. 3. XRD patterns for the $FeSe_{1.2}$ precursor and powder inside the FeSe wire annealed at 400-1000 °C. The symbols of "H" and "T" indicate the peaks of the hexagonal phase and the tetragonal phase, respectively. For $T_a$ = 1000 °C, a small peak of pure Fe was observed.

Fig. 4. $T_a$ dependence of existence ratio of the hexagonal phase estimated from intensity of those first peaks.

Fig. 5. Temperature dependence of total resistivity from 50 to 2 K for $T_a$ = 400 - 1000 °C.

Fig. 6. Annealing temperature dependence of $T_c^{onset}$ and $T_c^{offset}$.

Figure 1.

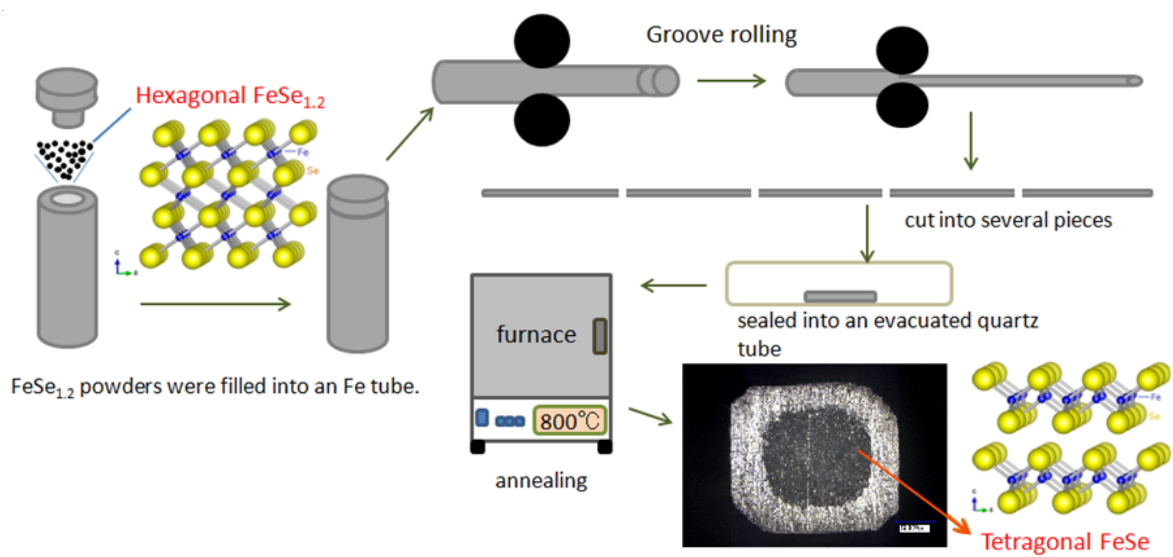

Figure 2.

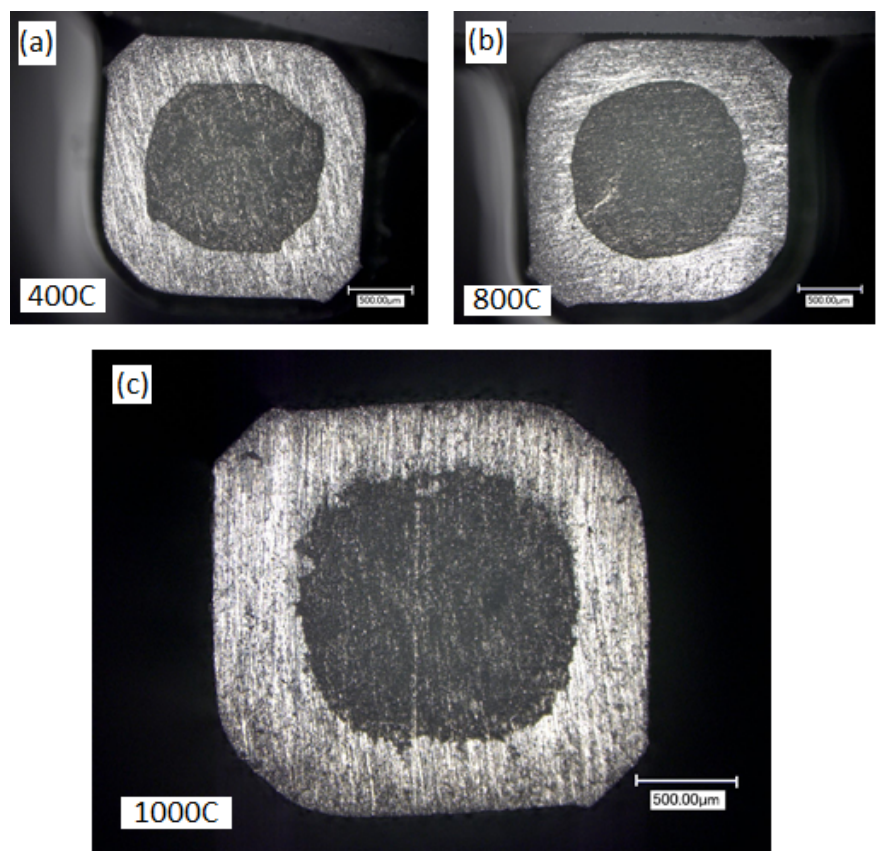

Figure 3.

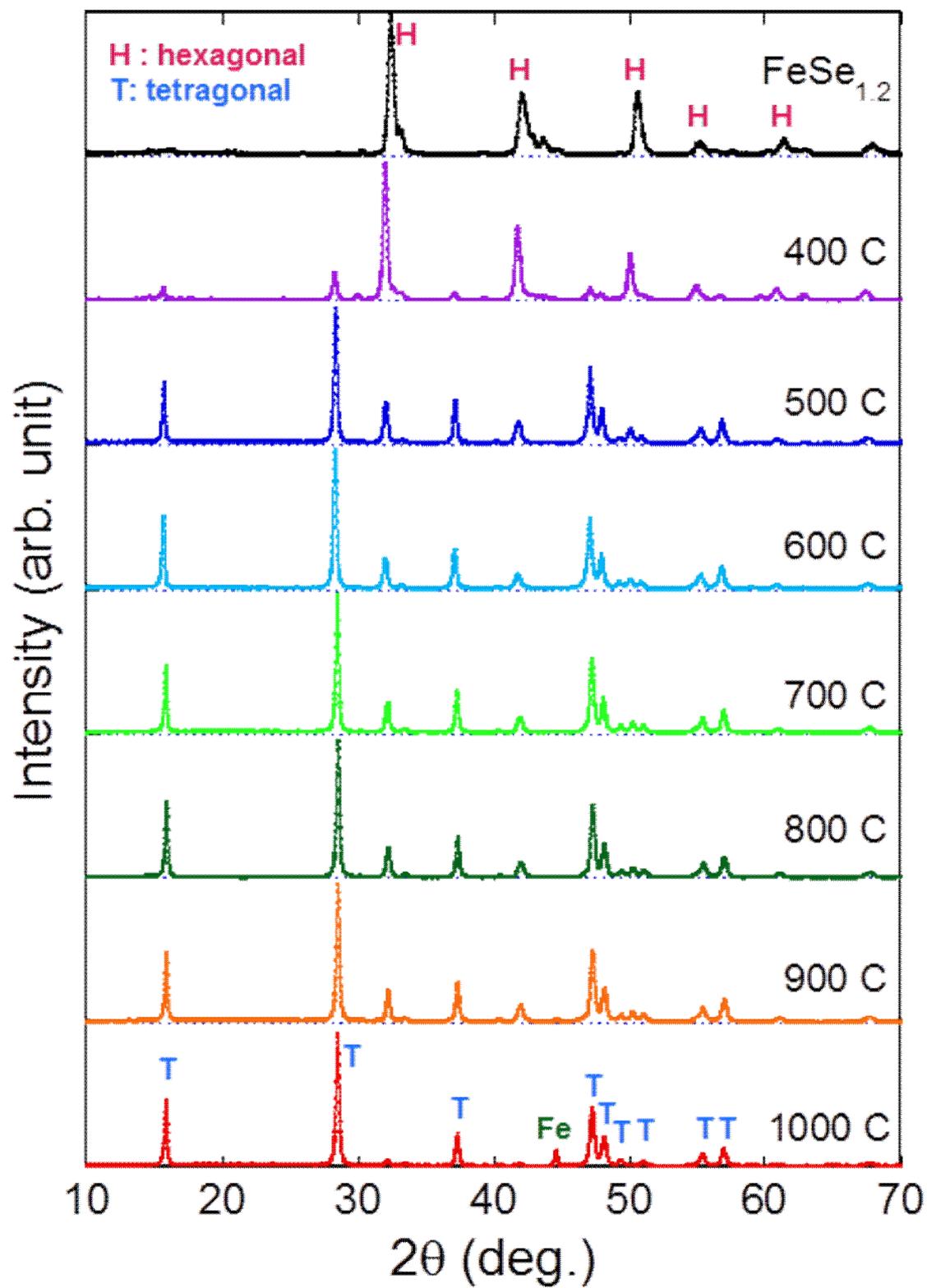

Figure 4.

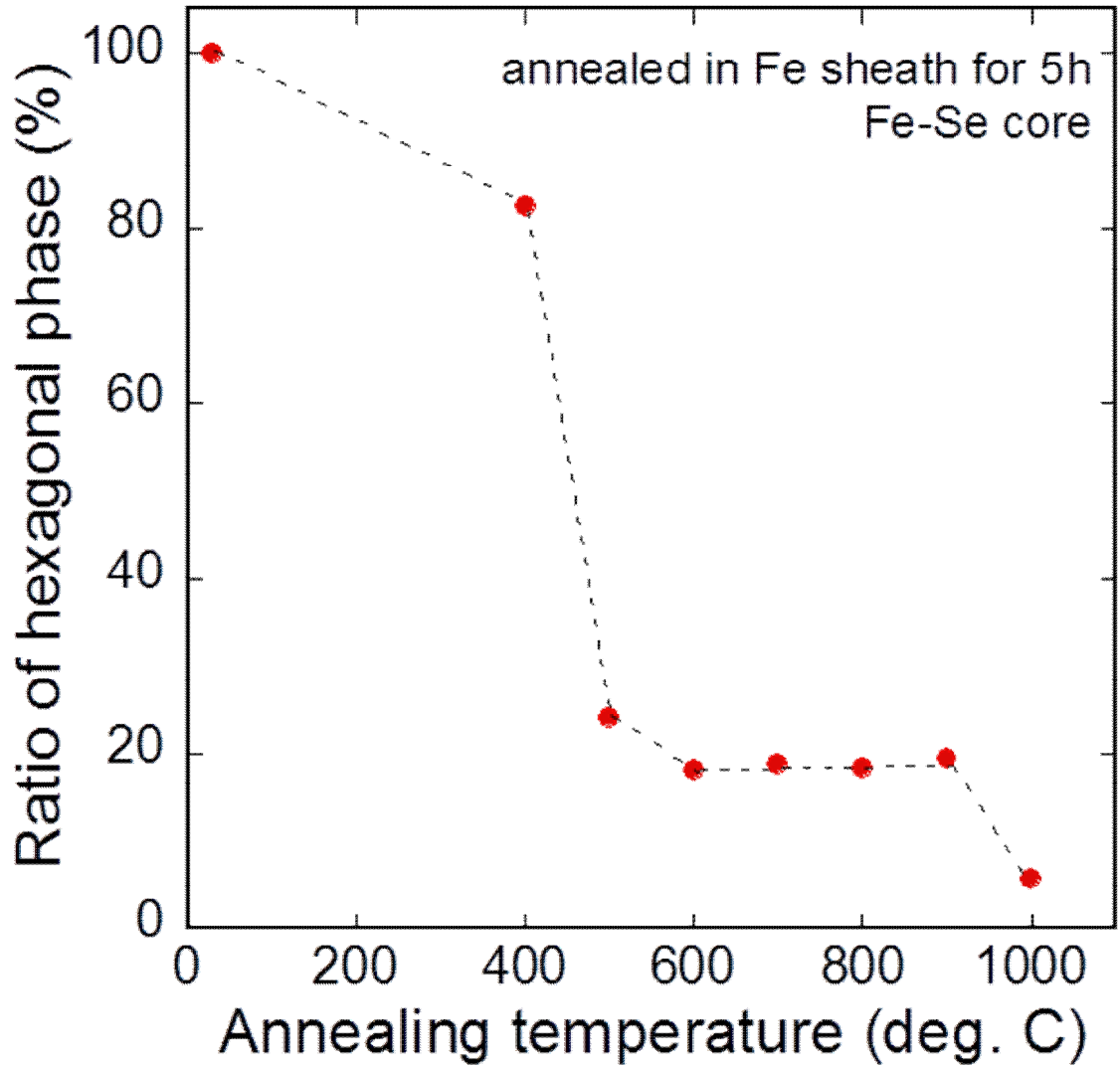

Figure 5.

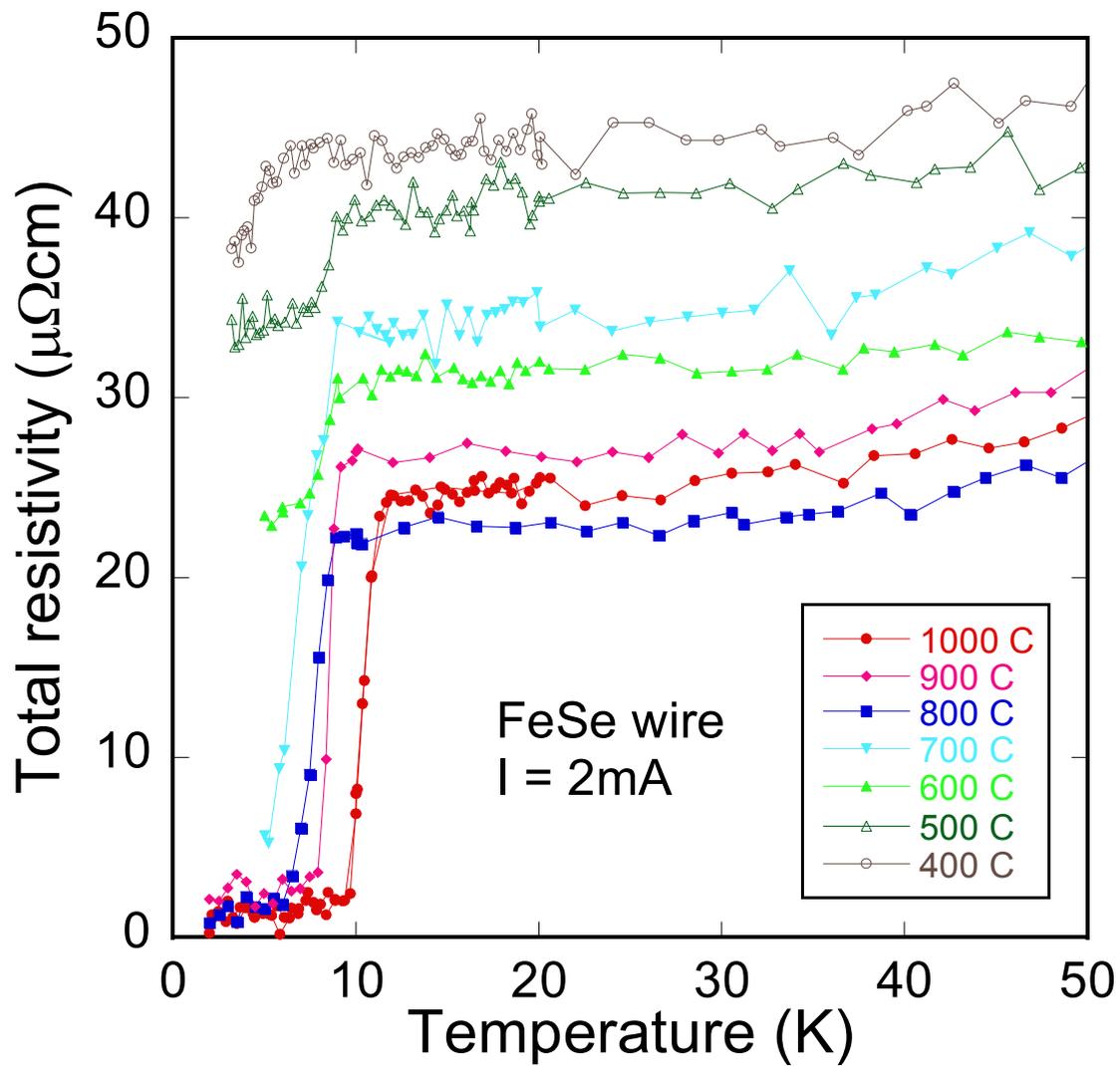

Figure 6.

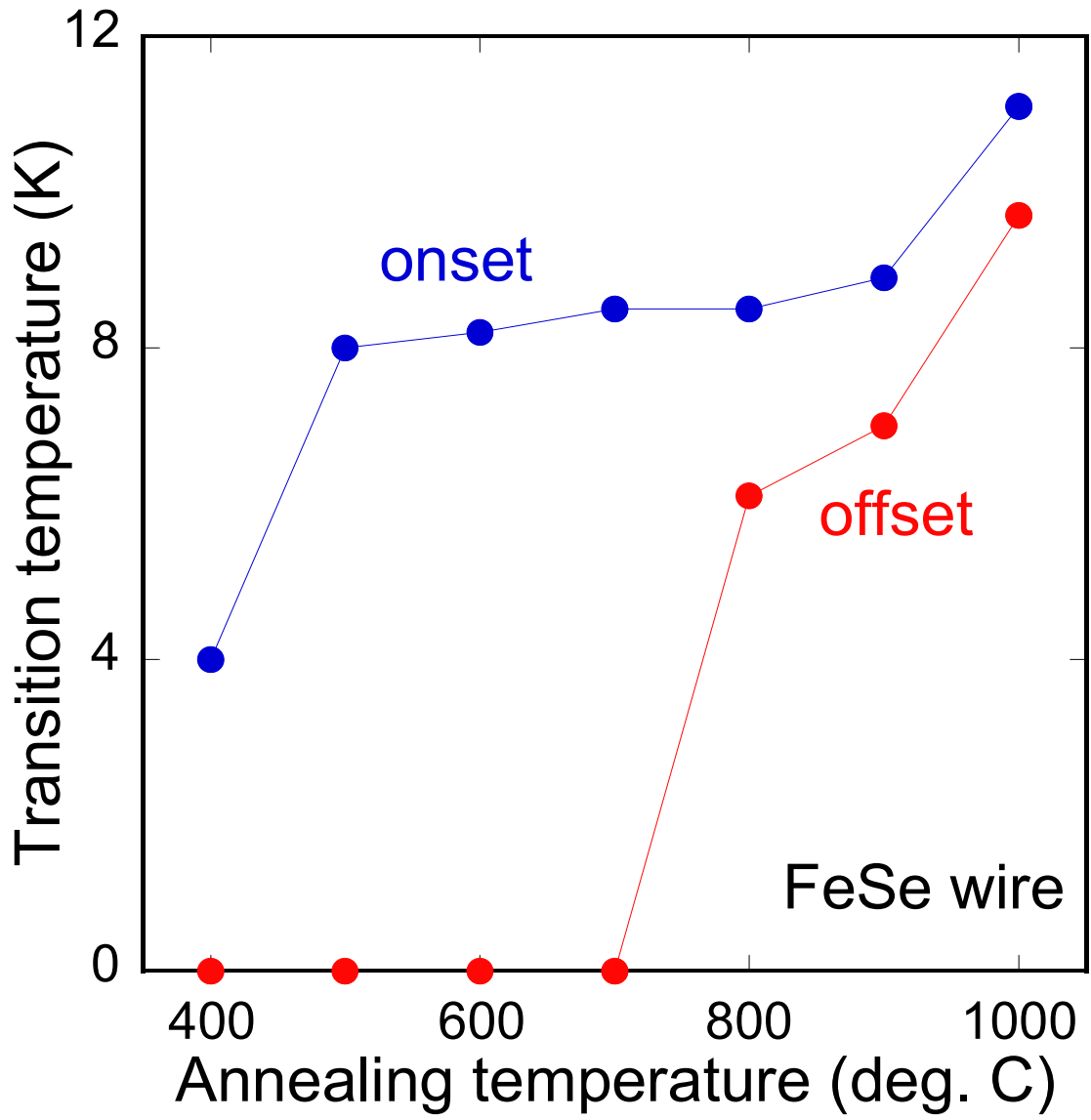